\documentclass{article}

\def\[{\left\lbrack}
\def\]{\right\rbrack}

\def\({\left(}
\def\){\right)}
\def\ih{\'\i}

\label{} 
\begin{document}
\date{}
\title{The Gauge Unfixing Formalism and the Solutions of the Dirac Bracket Commutators}
\author{Jorge Ananias Neto\\
Departamento de F\ih sica, ICE, \\ Universidade Federal de Juiz de Fora, 36036-900,\\ Juiz de Fora, MG, Brazil }

\maketitle

\begin{abstract}
We propose a systematic procedure that solves  the Dirac bracket commutators. The method is based on the Gauge 
Unfixing formalism, a procedure that converts second class systems into first class ones without the enlargement
of the original phase space variables. We verify that the gauge invariant variables satisfy the Dirac bracket 
when we strongly impose the discarded second class constraint. Thus, we can derive physical operators that 
satisfy the Dirac commutators. In order to illustrate our procedure, three 
second class constrained systems are considered. Firstly, the free particle on the two dimensional sphere is treated. The second case considered is the noncommutative free particle and the third is the doubly special relativity particle. 
\end{abstract}
\newpage

\section{Introduction}
The quantization of a dynamical system with second class constraints is usually performed by using the method proposed 
by Dirac, Bergman and coworkers\cite{Dirac}. The constraints are classified as primary and secondary ones. Secondary 
constraints are obtained from the condition that primary constraints are conserved in time. We must repeat the 
condition that requires vanishing time derivative of secondary constraints until all independent constraints are obtained.
If the whole second class constraints are established, then the so called Dirac bracket can be defined. In the case where there are two total second class constraints, the Dirac Bracket for the canonical variables $A$ and $B$ is given 
by\footnote{In the case where there is more than two second class constraints, the Dirac bracket for the variables $A$ and $B$ 
is given by $ \{A,B \}_{DB}= \{A,B \}- \{A,T_\alpha\}C^{-1}_{\alpha \beta} \{T_\beta,B\}$ where $T_\alpha$ and 
$T_\beta$ are the second class constraints and the matrix elements $ C_{\alpha\beta} $ is defined by $ C_{\alpha\beta} 
\equiv \{T_\alpha,T_\beta\}$.}

\begin{equation}
\label{db}
\{A,B \}_{DB}= \{A,B \}+ \frac{1}{ \{T_1,T_2\} }  \{A,T_1\} \{T_2,B\} - \{A,T_2\}\{T_1,B\},
\end{equation}
where $T_1$ and $T_2$ are the second class constraints.
The quantum mechanics commutators are given by the replacement $\left\lbrace \right\rbrace _{DB} \rightarrow  i \hbar
 \[,\]$. From the particular expressions of the Dirac bracket (DB) commutators, we can derive 
 the physical operators of a specific theory. These operators together with the physical states dictate the rules that govern the quantum system with constraints. However, the commutators are, in general, coordinate dependent
 and there are ordering problems. These facts make the task of obtaining the physical operators a very difficult one. Consequently, in principle, there is no general solution for the DB commutators. In this paper we develop, under certain conditions, a systematic procedure that derives physical operators in coordinates space that satisfy the DB commutators, at first,  for constrained systems with two second class constraints. 
 For this purpose, we use the Gauge Unfixing formalism (GU), a method that converts second class systems into first 
 class ones without the extension of the original phase space variables\cite{MR,Vyt}. We will see that, after converting the second class variables into first class ones and substituting these gauge invariant variables by operators, we obtain solutions of the DB commutators.  Thus, we can employ the idea of the GU formalism in order to establish a systematic procedure that solves the DB 
 commutators, . Aiming a logical presentation of the present work, this paper is organized as follows. In 
 Section 2 we give a short review of the GU formalism. In Section 3 we define a procedure, called Improved GU formalism, 
 where we obtain the gauge invariant variables. In Section 4 we verify that our gauge invariant variables satisfy the 
 Dirac bracket. In Sections 5, 6 and 7 we apply our formalism in the free particle constrained to the two dimensional 
 sphere \cite{JAN1}, in the noncommutative particle mechanics\cite{NCP} and the constrained doubly special relativity 
 particle\cite{DSR}, respectively. In Section 8 we make our concluding remarks.

\section{A Short Review of the Gauge Unfixing Formalism}

Let us consider a constrained system described by the second class Hamiltonian $H$ and two second class constraints 
$T_1$ and $T_2$. The basic idea of the GU formalism\cite{Vyt} is to select one of the two second class constraints to be the 
gauge symmetry generator. As example, if we choose $T_1$ as the first class constraint then the second class constraint $T_2$ will be discarded. The second class Hamiltonian 
must be modified in order to satisfy a first class algebra. The gauge invariant Hamiltonian is constructed from 
a series in powers of $T_2$

\begin{eqnarray}
\label{gh}
\tilde{H}=H - \frac{1}{\delta T_2}\{H,T_1\}\; T_2 + \frac{1}{2!} \frac{1}{\delta T_2}
\{ \{H,T_1\},T_1\} \; (T_2)^2\nonumber \\-
\frac{1}{3!}\frac{1}{\delta T_2} \{\{\{ H,T_1\},T_1\},T_1\}\; (T_2)^3+\ldots,
\end{eqnarray}
where $\delta T_2 \equiv \{T_2,T_1\}$. From Eq.(\ref{gh}), we can show that $\{\tilde{H},T_1\}=0$, and $T_1$ must satisfy a first class algebra, 
$\{T_1,T_1\}=0$. 

\section{The Improved Gauge Unfixing Formalism}
Let us start with the original phase space variables written as

\begin{equation}
F=(q_i,p_i),
\end{equation}
where $\, F \,$ can describe a particle or field model. As we have seen in Section 2, the usual GU formalism embeds directly the second class Hamiltonian. Thus, our strategy is to construct a gauge invariant function  $\tilde{A}\,$  from the second class function  $A\,$  by gauging the original phase space variables, using for this the idea of the GU formalism. Denoting the first class variables by

\begin{equation}
\tilde{F}=(\tilde{q_i},\tilde{p_i}),
\end{equation}
we determine the first class function $\tilde{F}$ in terms of the original phase space variables by employing the variational condition 

\begin{equation}
\label{vc}
\delta\tilde{F}=\epsilon \{\tilde{F},\tilde{T}\}=0,
\end{equation}
where $\tilde{T}$ is the second class constraint chosen to be the gauge symmetry generator and $\epsilon$ is an infinitesimal parameter. Any function of $\tilde{F}\,$ will be gauge invariant since

\begin{equation}
\{\tilde{A}(\tilde{F}),\tilde{T}\}=\{\tilde{F},\tilde{T}\} \frac {\partial\tilde{A}} {\partial\tilde{F}}=0,
\end{equation}
where

\begin{equation}
\{\tilde{F},\tilde{T}\} \frac {\partial\tilde{A}} {\partial\tilde{F}}\equiv\{\tilde{q}_i,\tilde{T}\}\frac {\partial\tilde{A}}{\partial\tilde{q}_i}+ \{\tilde{p}_i,\tilde{T}\}\frac {\partial\tilde{A}}{\partial\tilde{p}_i}.
\end{equation}
Consequently, we can obtain a gauge invariant function from the replacement of

\begin{equation}
\label{sub}
A(F)\Rightarrow A(\tilde{F})=\tilde{A}(\tilde{F}).
\end{equation}
The gauge invariant phase space variables $\tilde{F}$ are constructed by the series in powers of $T_2$

\begin{equation}
\label{ff}
\tilde{F}=F+ \sum_{n=1}^{\infty} c_n\,T_2^n=F+c_1\,T_2+c_2\,T_2^2+\ldots,
\end{equation}
where this series has an important boundary condition that is

\begin{equation}
\label{bon}
\tilde{F}(T_2=0)=F.
\end{equation}
The condition above and the relation (\ref{sub}) show that when we impose the discarded  constraint $T_2$ equal to zero, we reobtain the original second class system. Therefore, the relations (\ref{sub}) and (\ref{bon})  guarantee the equivalence between our first class model and the initial second class system.
The coefficients $c_n$ in the relation (\ref{ff}) are then determined by the variational condition, Eq.(\ref{vc}). The general equation for $c_n$ is

\begin{equation}
\delta\tilde{F}=\delta F +\sum_{n=1}^\infty\, (\delta c_n\, T_2^n+ n\, c_n\,T_2^{(n-1)}\delta T_2)=0,
\end{equation} 
where

\begin{eqnarray}
\delta F &=&\epsilon \{F,\tilde{T}\},\\
\delta c_n &=&\epsilon \{c_n,\tilde{T}\},\\
\label{res}
\delta T_2&=&\epsilon \{T_2,\tilde{T}\} ,
\end{eqnarray}
where $\, \epsilon \, $ is an infinitesimal parameter. Then, for the linear correction term $(n=1)$, we have

\begin{equation}
\label{c_1}
\delta F+ c_1\,\delta T_2=0 \; \Rightarrow\; c_1=-\frac{\delta F}{\delta T_2}.
\end{equation}
For the quadratic correction term (n=2), we get

\begin{equation}
\label{c_2}
\delta c_1+2c_2\,\delta T_2=0 \; \Rightarrow\; c_2=-\frac{1} {2}\frac{\delta c_1}{\delta T_2}.
\end{equation}
For $n\geq 2$, the general relation is

\begin{eqnarray}
\label{c_n}
\delta c_n+(n+1)c_{n+1}\,\delta T_2=0\;\Rightarrow\; c_{(n+1)}=-\frac{1}{(n+1)}\frac{\delta c_n}{\delta T_2}.
\end{eqnarray}
Using the relations (\ref{c_1}), (\ref{c_2}) and (\ref{c_n}) in Eq.(\ref{ff}) we obtain the series which determines $\tilde{F}$

\begin{equation}
\label{series}
\tilde{F}= F - \frac{\delta F}{\delta T_2}\,T_2
+\frac{1}{2!} \frac{\delta(\frac{\delta F}{\delta T_2})}{\delta T_2}\,(T_2)^2
-\frac{1}{3!} \frac{\delta\delta(\frac{\delta F}{\delta T_2})}{\delta T_2}\,(T_2)^3 + \ldots \,.
\end{equation}
We can verify that our gauge invariant variable, $\tilde{F}$, satisfies the condition $\delta \tilde{F}=\epsilon \{\tilde{F},\tilde{T}\}=0$.

\section{The GU Variables as a Route to Find Solutions of the Dirac Bracket Commutators}

Evaluating the Poisson bracket between the two gauge invariant variables defined by the formula (\ref{series}) and taking the limit $T_2 \rightarrow 0$,  we get

\begin{eqnarray}
\label{pd}
\{ \tilde{F},\tilde{G} \}_{T_2 \rightarrow 0} = 
\{F,G\}+  \frac{1}{\{T_1,T_2\}}\, \{F,T_1\} \{T_2,G\}
- \{F,T_2\}\{T_1,G\}\nonumber\\
= \{F,G\}_{DB},
\end{eqnarray}
where $\{F,G\}_{DB}$ is the Dirac bracket defined in Eq.(\ref{db}). Thus, the Dirac bracket algebra can be reproduced by the Poisson bracket algebra between the gauge invariant variables\footnote{This same result occurs in the BFT formalism\cite{Hong}.} in the weak sense\cite{Dirac}. Then, explicit solutions of the DB commutators can be obtained by replacing\footnote{We have used the usual correspondence principle\cite{Dirac} between the classical mechanics and the quantum mechanics.} the classical variables $x_i$ and $p_i$ by the operators $\hat{x}_i=x_i$ and $\hat{p}_i=-i\hbar \frac{\partial}{\partial x^i}$ in the gauge invariant variables, Eq.(\ref{series}). Therefore, given the phase space variables of a specific second class constrained system, we can derive solutions of the quantum mechanics commutators by developing the Gauge Unfixing variables and then promoting the classical variables to the operators. Here, we would like to comment that the operator solutions present, in general, ordering problem. At first, we can solve this difficulty by using the Weyl ordering operator prescription\cite{Weyl}. For example, if we have two noncommuting operators $\hat{A}$ and $\hat{B}$, then we must replace the product $\hat{A}$ $\hat{B}$ by the symmetrization procedure 

\begin{eqnarray}
\hat{A} \hat{B} \rightarrow \frac{1}{2}(\hat{A} \hat{B}+\hat{B}\hat{A}).
\end{eqnarray}

\section{The Free Particle Constrained on the Two Dimensional Sphere}

The dynamic of a particle in the two dimensional sphere has the primary constraint given by

\begin{equation}
\label{pri}
\phi_1 \equiv x^2-R^2 \approx 0,
\end{equation}
where $x^2\equiv x_ix_i$ and $R$ is the radius of the sphere. With the expression of the classical Hamiltonian

\begin{equation}
H={1\over 2} \pi_i \pi_i \,\,,
\end{equation}
we obtain the secondary constraint

\begin{equation}
\label{sec}
\phi_2\equiv x\pi \approx 0,
\end{equation}
where $x\pi\equiv x_i\pi_i$. This constraint expresses the fact that motion on the surface of a sphere has no 
radial component. We observe that no further constraints are generated via this iterative procedure. $\phi_1$ 
and $\phi_2$ are the total second class constraints of the model. From the expressions of the constraints, 
Eqs.(\ref{pri}) and (\ref{sec}), and using the Dirac bracket formula, Eq.(\ref{db}), we obtain the algebra 
of the canonical variables

\begin{eqnarray}
\{ x_i,x_j \}_{DB} &=& 0,\\
\{ x_j,\pi_k \}_{DB} & =& \delta_{jk} -\frac{ x_j x_k}{R^2},\\
\{\pi_j,\pi_k \}_{DB} & = & \frac{1}{R^2} (x_k\pi_j-x_j\pi_k ).
\end{eqnarray}
Therefore the quantum commutators are

\begin{eqnarray}
\label{cx}
\[ x_i,x_j \]&=& 0,\\
\label{cxp}
\[ x_j,\pi_k \] &=& i\hbar (\delta_{jk} - \frac{x_j x_k}{R^2} ),\\
\label{cpp}
\[ \pi_j,\pi_k \]& =& i\hbar(\frac{1}{R^2} (x_k\pi_j -x_j\pi_k )).
\end{eqnarray}
Thus, our objective is to find solutions for the position and momentum operators that satisfy the 
commutators, Eqs. (\ref{cx}),(\ref{cxp}) and (\ref{cpp}). For this purpose, we will apply the 
GU formalism. The first step is to select the symmetry gauge generator. We choose

\begin{equation}
\label{fi1}
\tilde{\phi}=\phi_1= x^2 - R^2.
\end{equation}
The second class constraint $\phi_2=x\pi$ will be discarded. The infinitesimal gauge transformations generated by the symmetry generator $\tilde{\phi}$, Eq.(\ref{fi1}), are

\begin{eqnarray}
\delta x_i &=& \epsilon \{x_i,\tilde{\phi} \} = 0.\\
\delta \pi_i &=& \epsilon \{\pi_i, \tilde{\phi} \} = - 2 \,\epsilon \,x_i,\\
\delta \phi_2 &=& \epsilon \{\phi_2,\tilde{\phi}\} = -2\, \epsilon \, x^2.
\end{eqnarray}
The gauge invariant position $\tilde{x}_i$ is developed by the series in powers of $\phi_2$

\begin{equation}
\tilde{x}_i=x_i+ b_1\,\phi_2+ b_2\,(\phi_2)^2 +
 ... + b_n\,(\phi_2)^n.
 \end{equation}
From the invariance condition $\delta \tilde{x}_i=0$, we can compute all the correction terms $b_n$. For the linear correction term in order of $\phi_2$, we get

\begin{eqnarray}
\delta x_i + b_1 \delta \phi_2 = 0 \nonumber\\
\Rightarrow b_1 = 0.
\end{eqnarray}
Due to the fact that $b_1=0$, all the correction terms $b_n$ are null. Therefore the gauge invariant position $\tilde{x}_i$ is

\begin{equation}
\label{x1}
\tilde{x}_i=x_i.
\end{equation}
The gauge invariant momentum $\tilde{\pi}_i$ is built by the series in powers of $\phi_2$

\begin{equation}
\tilde{\pi}_i= \pi_i+ c_1\,\phi_2+ c_2\,(\phi_2)^2 +
 ... + c_n\,(\phi_2)^n .
\end{equation}
From the invariance condition $\delta\tilde{\pi}_i=0$, we can calculate all the correction terms $c_n$. For the linear correction in order of $\phi_2$, we get
\begin{eqnarray}
\delta \pi_i + c_1 \delta \phi_2 = 0 \nonumber\\
\Rightarrow c_1 = - \frac{x_i}{x^2}.
\end{eqnarray}
For the quadratic term, we obtain $c_2=0$, since $\delta c_1=\epsilon \{c_1,\tilde{\phi}\}=0$. Due to this, all the correction terms $c_n$ with $n\geq 2$ are null. Then, the gauge invariant momentum $\tilde{\pi}_i$ is

\begin{equation}
\label{p1}
\tilde{\pi}_i = \pi_i - \frac{x_i}{x^2}\, x\pi.
\end{equation}
Then, using Eqs.(\ref{x1}) and (\ref{p1}), we find the operators solutions of the DB commutators, Eqs.(\ref{cx}),(\ref{cxp}) and (\ref{cpp}), as

\begin{eqnarray}
\label{xs}
\hat{x}_i&=& x_i,\\
\label{ps}
\hat{\pi}_i&=& -i\hbar \[ \partial_i - \frac{x_i x_j\partial_j}{x^2} \],
\end{eqnarray}
where we replace $\pi_i$ by $-i\hbar\,\partial_i$. These solutions can be used to build the Hamiltonian operator, $H=\frac{1}{2}\,\hat{\pi}_i\hat{\pi}_i$. It is important to observe the  ordering problem that appears in Equation (\ref{ps}). However, we can solve this problem by using the Weyl ordering operator prescription (\cite{Weyl}.

\section{Noncommutative Particle Mechanics}
The Lagrangian of the relativistic free particle is
\begin{eqnarray}
L= - m\,\sqrt{\dot{x_\mu} \dot{x^\mu}},
\end{eqnarray}
with $x^\mu, \mu=0,1,...,d$ and the dot means differentiation with time $\tau$. The conjugate momentum 

\begin{equation}
p_\mu\equiv\frac{\partial L}{\partial \dot{x}^\mu} = \frac{m\,\dot{x}_\mu}{\sqrt{(\dot{x}^\nu)^2}}
\end{equation}
leads to the constraint that is the mass shell condition

\begin{equation}
\label{ncp1}
\phi_1\equiv p^2-m^2=p_0p_0-p_ip_i-m^2\approx 0,
\end{equation}
being the metric $g_{\mu\nu}=diag(1,-1,...,-1)$. Due to the reparametrization invariance, the Hamiltonian vanish

\begin{equation}
H = p\,\dot{x} - L = 0.
\end{equation}
The gauge symmetry can be fixed by imposing a gauge condition. In the noncommutative particle mechanics (NCP) we choose\cite{NCP}

\begin{equation}
\label{ncp2}
\phi_2 \equiv x_0+\theta_{0i}p_i-\tau\approx 0, \;\; i=1,2,...,d,
\end{equation}
where $\theta_{0i}$ is a constant. If we make the parameter $\theta_{0i}$ equal to zero, we recover the commutative relativistic particle model. The constraints (\ref{ncp1}) and (\ref{ncp2}) form a second class set with

\begin{equation}
\{\phi_1,\phi_2\}=-2 p_0.
\end{equation}
Using the Dirac brackets (\ref{db}), we obtain

\begin{eqnarray}
\{x_0,x_i\}_{DB}&=& \theta_{0i},\\
\{x_i,x_j\}_{DB}&=&\frac{1}{p^0} (\theta_{0i} p_j - \theta_{0j} p_i),\\
\{x_i,p_0\}_{DB}&=&\frac{p_i}{p^0},\\
\{x_i,p_j\}_{DB}&=&\delta_{ij}.
\end{eqnarray}
Therefore the quantum commutators are

\begin{eqnarray}
\label{ncpc0x}
\[ x_0,x_i \]&=& i\hbar\,\theta_{0i},\\
\label{ncpcxj}
\[ x_i,x_j \] &=& i\hbar\,(\frac{1}{p^0} (\theta_{0i} p_j - \theta_{0j} p_i)) ,\\
\label{ncpcxp0}
\[ x_i,p_0 \]& =& i\hbar\,\frac{p_i}{p^0},\\
\label{ncpxipj}
\[x_i,p_j\]&=&  i\hbar\,\delta_{ij}.
\end{eqnarray}
In order to find solutions for the position and momentum operators that satisfy the relations, Eqs.(\ref{ncpc0x}),(\ref{ncpcxj}),(\ref{ncpcxp0}) and (\ref{ncpxipj}), we will apply the 
GU formalism. The first step is to select the symmetry gauge generator. We choose

\begin{equation}
\label{fit}
\tilde{\phi}\equiv\phi_1= p_0p_0-p_ip_i-m^2.
\end{equation}
The second class constraint, $\phi_2=x_0+\theta_{0i}p_i-\tau$, will be discarded. The infinitesimal gauge transformations generated by the symmetry generator $\tilde{\phi}$, Eq.(\ref{fit}), are

\begin{eqnarray}
\delta x_0 &=& \epsilon \{x_0,\tilde{\phi}\}=2 \epsilon p_0,\\
\delta x_i &=& \epsilon \{x_i,\tilde{\phi}\}=-2\epsilon p_i,\\
\delta p_0 &=& \epsilon \{p_0, \tilde{\phi}\}=0,\\
\delta p_i &=& \epsilon \{p_i, \tilde{\phi}\}=0,\\
\delta \phi_2 &=& \epsilon \{\phi_2,\tilde{\phi}\}= 2 \epsilon p_0 .
\end{eqnarray}
The gauge invariant variable $\tilde{x}_0$ is constructed by the series in powers of $\phi_2$

\begin{equation}
\tilde{x}_0= x_0+ d_1\,\phi_2+ d_2\,(\phi_2)^2 +
 ... + d_n\,(\phi_2)^n.
\end{equation}
From the invariance condition $\delta \tilde{x}_0=0$, we can compute all the correction terms $d_n$. For the linear correction term in order of $\phi_2$, we get

\begin{eqnarray}
\delta x_0 + d_1 \delta \phi_2 = 0 \nonumber\\
\Rightarrow d_1 = -1.
\end{eqnarray}
For the quadratic term, we obtain $d_2=0$, since $\delta d_1=\epsilon \{d_1,\tilde{\phi}\}=0$. Due to this, all the correction terms $d_n$ with $n\geq 2$ are null. Therefore the gauge invariant variable $\tilde{x}_0$ is

\begin{equation}
\label{ncpx0}
\tilde{x}_0 = x_0-(x_0+\theta_{0i} p_i-\tau)=-\theta_{0i} p_i+\tau.
\end{equation}
The gauge invariant position $\tilde{x}_i$ is constructed by the series in powers of $\phi_2$

\begin{equation}
\tilde{x}_i= x_i + e_1\,\phi_2+ e_2\,(\phi_2)^2 +
 ... + e_n\,(\phi_2)^n.
\end{equation}
From the invariance condition $\delta \tilde{x}_i=0$, we can calculate all the correction terms $e_n$. For the linear correction term in order of $\phi_2$, we get

\begin{eqnarray}
\delta x_i + e_1 \delta \phi_2 = 0 \nonumber\\
\Rightarrow e_1 = \frac{p_i}{p_0}.
\end{eqnarray}
For the quadratic term, we obtain $e_2=0$, since $\delta e_1=\epsilon \{e_1,\tilde{\phi}\}=0$. Due to this, all the correction terms $e_n$ with $n\geq 2$ are null. Then, the gauge invariant position $\tilde{x}_i$ is

\begin{equation}
\label{ncpxi}
\tilde{x}_i = x_i + \frac{p_i}{p^0} (x_0+\theta_{0j} p_j - \tau).
\end{equation}
The gauge invariant variable $\tilde{p}_0$ is constructed as

\begin{equation}
\tilde{p}_0= p_0+ f_1\,\phi_2+ f_2\,(\phi_2)^2 +
 ... + f_n\,(\phi_2)^n.
\end{equation}
From the invariance condition $\delta \tilde{p}_0=0$, we can compute all the correction terms $f_n$. For the linear correction term in order of $\phi_2$, we obtain

\begin{eqnarray}
\delta p_0 + f_1 \delta \phi_2 = 0 \nonumber\\
\Rightarrow f_1 = 0.
\end{eqnarray}
Due to the fact that $f_1=0$, all the correction terms $f_n$ are null. Then, the gauge invariant variable $\tilde{p}_0$ is

\begin{equation}
\label{ncpp0}
\tilde{p}_0=p_0.
\end{equation}
The gauge invariant momentum $\tilde{p}_i$ is given by

\begin{equation}
\tilde{p}_i=p_i+ g_1\,\phi_2+ g_2\,(\phi_2)^2 +
 ... + g_n\,(\phi_2)^n.
 \end{equation}
From the invariance condition $\delta \tilde{p}_i=0$, we can calculate all the correction terms $g_n$. For the linear correction term in order of $\phi_2$, we get

\begin{eqnarray}
\delta p_i + g_1 \delta \phi_2 = 0 \nonumber\\
\Rightarrow g_1 = 0.
\end{eqnarray}
Due to the fact that $g_1=0$, all the correction terms $g_n$ are null. Therefore, the gauge invariant momentum $\tilde{p}_i$ is

\begin{equation}
\label{ncppi}
\tilde{p}_i=p_i.
\end{equation}
Using Eqs.(\ref{ncpx0}), (\ref{ncpxi}), (\ref{ncpp0}) and (\ref{ncppi}), we find the operators solutions of the DB commutators, Eqs.(\ref{ncpc0x}), (\ref{ncpcxj}), (\ref{ncpcxp0}) and (\ref{ncpxipj}), as

\begin{eqnarray}
\label{xoop}
\hat{x}_0&=& -\theta_{0j} p_j+\tau,\\
\label{xiop}
\hat{x}_i &=& x_i + \frac{p_i}{p^0} [x_0+\theta_{0j} p_j - \tau],\\
\label{poop}
\hat{p}_0&=& p_0 ,\\
\label{piop}
\hat{p}_i&=& p_i.
\end{eqnarray}
Here $p_0 \equiv -i\hbar\,\partial_0\,$ and $p_i \equiv -i\hbar\,\partial_i\,$. Equations(\ref{xoop}),(\ref{xiop}), (\ref{poop}) and (\ref{piop}) are new results obtained with the aid of the GU formalism. In the Eq.(\ref{xiop}) we can move the operator $p_0$ from the denominator for the numerator by performing a binomial expansion in powers of $(p_0-1)$. Then the operator $\hat{x}_i$ can be written as

\begin{eqnarray}
\label{xip0}
\hat{x}_i=x_i + p_i (x_0+\theta_{0j} p_j - \tau)[1-(p_0-1)+(p_0-1)^2- \dots].
\end{eqnarray}
We can observe that the Eq.(\ref{xiop}) is equivalent to the series, Eq.(\ref{xip0}), with higher order terms in $p_0$.
As we have mentioned in the previous section, the Weyl ordering operator prescription can be used in order to solve the ordering problem that appears in Eq.(\ref{xip0}).

\section{Doubly Special Relativity Particle}
Motivated by the ideas of quantum gravity, several authors\cite{DSR} have proposed a model called Doubly Special Relativity 
Particle (DSR) which is similar to the Special Theory of Relativity. The Special Theory of Relativity has one 
observer independent 
scale that is the velocity of light $\,c\,$. DSR theory have two observer independent scales which are the usual 
velocity of light and a length scale. This model can be described by the Lagrangian  proposed 
by Ghosh\cite{Ghosh} 
\begin{eqnarray}
\label{ldsr}
L = \frac{m k}{\sqrt{k^2-m^2}}\, [ \,g_{\mu\nu} \dot{x}^\mu\dot{x}^\nu + \frac{m^2}{k^2-m^2}(g_{\mu\nu} \dot{x}^\mu 
\eta^\nu)^2 ]^{1/2}\nonumber\\ - \frac{m^2 k}{k^2 - m^2} \, g_{\mu\nu} \dot{x}^\mu \eta^\nu,
\end{eqnarray}
where $\eta^0 = 1, \vec{\eta} = 0 $ and $\,k\,$ is a parameter related to the Planck mass. The conjugate momentum 

\begin{eqnarray}
p_\mu = \frac{\partial L}{\partial \dot{x^\mu}}
=\frac{m k}{\sqrt{k^2-m^2}}\,\frac{\dot{x_\mu}+ \frac{m^2}{k^2-m^2}(g_{\mu\nu}\dot{x}^\mu 
\eta^\nu)\eta_\mu}{\Lambda}\nonumber\\
-\,\frac{m^2\,k}{k^2-m^2}\,\eta_\mu,
\end{eqnarray}
where $\Lambda \equiv \sqrt{ g_{\mu\nu} \dot{x}^\mu\dot{x}^\nu + \frac{m^2}{k^2-m^2}(g_{\mu\nu} \dot{x}^\mu \eta^\nu)^2} \,$,
leads to the Magueijo-Smolin (MS) dispersion relation\cite{MS}

\begin{equation}
\label{MSR}
p^2=m^2(1-\frac{\eta p}{k})^2,
\end{equation}
where we have adopted the notation $AB\equiv g_{\mu\nu} A^\mu B^\nu$ with the metric 
$g_{\mu\nu}=diag(1,-1,...,-1)$. We can observe that making the limit $k\rightarrow\infty$ in the Eq.(\ref{MSR}), we recover the usual relation $p^2=m^2$. Due to the $\tau$-reparametrization invariance of the Lagrangian (\ref{ldsr}), the Hamiltonian vanish

\begin{equation}
H = p\dot{x} - L = 0.
\end{equation}
Then, we have a first class system with the MS dispersion relation, Eq.(\ref{MSR}), being the first class constraint
\begin{eqnarray}
\phi_1\equiv p^2-m^2(1-\frac{\eta p}{k})^2\approx 0.
\end{eqnarray}
Thus, if we choose the gauge condition as\cite{Ghosh} 

\begin{equation}
\phi_2 \equiv xp \approx 0,
\end{equation}
we have a second class constrained system with

\begin{equation}
\left\lbrace  \phi_1,\phi_2 \right\rbrace = - 2 m^2 (1-\frac{\eta p}{k}).
\end{equation}
Using Eq.(\ref{db}) we obtain the Dirac brackets

\begin{eqnarray}
\{x_\mu,x_\nu\}_{DB} &=&-\frac{1}{k} ( x_\mu\eta_\nu-x_\nu\eta_\mu )
- \frac{1}{m^2(1-\frac{\eta p}{k})} (x_\mu p_\nu-x_\nu p_\mu),\\
\nonumber\\
\{x_\mu,p_\nu\}_{DB} &=& \delta_{\mu\nu} - \frac{1}{k} \eta_\mu p_\nu - \frac{p_\mu p_\nu} {m^2(1-\frac{\eta p}{k})},\\
\nonumber\\
\{p_\mu,p_\nu\}_{DB} &=&0.
\end{eqnarray}
Hence the quantum commutators are

\begin{eqnarray}
\label{xxdsr}
\[x_\mu,x_\nu\]&=& i\hbar \[ -\frac{1}{k} ( x_\mu\eta_\nu-x_\nu\eta_\mu )
- \frac{1}{m^2(1-\frac{\eta p}{k})} (x_\mu p_\nu-x_\nu p_\mu)\],\\
\nonumber\\
\label{xpdsr}
\[x_\mu,p_\nu\]&=& i\hbar\[\delta_{\mu\nu} - \frac{1}{k} \eta_\mu p_\nu - \frac{p_\mu p_\nu} {m^2(1-\frac{\eta p}{k})}\],\\
\nonumber\\
\label{ppdsr}
\[p_\mu,p_\nu\]&=&0.
\end{eqnarray}
In order to find solutions for the position and momentum operators that satisfy the complicated relations, Eqs.(\ref{xxdsr}),(\ref{xpdsr}) and (\ref{ppdsr}), we will use the GU formalism. We choose the symmetry gauge generator as

\begin{equation}
\label{fi1dsr}
\tilde{\phi}\equiv \phi_1= p^2-m^2(1-\frac{\eta p}{k})^2.
\end{equation}
The constraint $\phi_2=xp$ will be discarded. The infinitesimal gauge transformations generated by the symmetry generator $\tilde{\phi}$, Eq.(\ref{fi1dsr}), are

\begin{eqnarray}
\delta x_\mu &=& \epsilon \{x_\mu,\phi_1\}=2 \epsilon (p_\mu+\frac{m^2}{k} \eta_\mu (1-\frac{\eta p}{k})),\\
\delta p_\mu &=& \epsilon \{p_\mu, \phi_1\}=0,\\
\delta \phi_2 &=& \epsilon \{\phi_2,\phi_1\}= 2 m^2 \epsilon (1-\frac{\eta p}{k}) .
\end{eqnarray}
The gauge invariant variable $\tilde{x}_\mu$ is constructed by the series in powers of $\phi_2$

\begin{equation}
\tilde{x}_\mu= x_\mu+ h_1\,\phi_2+ h_2\,(\phi_2)^2 +
 ... + h_n\,(\phi_2)^n.
\end{equation}
From the invariance condition $\delta \tilde{x}_\mu=0$, we can compute all the correction terms $h_n$. For the linear correction term in order of $\phi_2$, we get

\begin{eqnarray}
\delta x_\mu + h_1 \delta \phi_2 = 0 \nonumber\\
\Rightarrow h_1 = - \[\frac{p_\mu}{m^2(1-\frac{\eta p}{k})} + \frac{\eta_\mu}{k}\].
\end{eqnarray}
For the quadratic term, we obtain $h_2=0$, since $\delta h_1=\epsilon \{h_1,\tilde{\phi\}}=0$. Due to this, all the correction terms $h_n$ with $n\geq 2$ are null. Therefore the gauge invariant variable $\tilde{x}_\mu$ is

\begin{eqnarray}
\label{dsrxmu}
\tilde{x}_\mu = x_\mu -  \[\frac{p_\mu}{m^2(1-\frac{\eta p}{k})} + \frac{\eta_\mu}{k}\]\, xp.
\end{eqnarray}
The gauge invariant variable $\tilde{p}_\mu$ is constructed by the series in powers of $\phi_2$

\begin{equation}
\tilde{p}_\mu=p_\mu+ i_1\,\phi_2+ i_2\,(\phi_2)^2 +
 ... + i_n\,(\phi_2)^n.
 \end{equation}
From the invariance condition $\delta \tilde{p}_\mu=0$, we can compute all the correction terms $i_n$. For the linear correction term in order of $\phi_2$, we get

\begin{eqnarray}
\delta p_\mu + i_1 \delta \phi_2 = 0 \nonumber\\
\Rightarrow i_1 = 0.
\end{eqnarray}
Due to the fact that $i_1=0$, all the correction terms $i_n$ are null. Then, the gauge invariant variable $\tilde{p}_\mu$ is

\begin{equation}
\label{dsrpmu}
\tilde{p}_\mu=p_\mu.
\end{equation}
Using Eqs.(\ref{dsrxmu}) and (\ref{dsrpmu}), we find the operators solutions of the DB commutators, Eqs.(\ref{xxdsr}), (\ref{xpdsr}) and (\ref{ppdsr}), as

\begin{eqnarray}
\label{ordxmu}
\hat{x}_\mu&=& x_\mu - \[\frac{\eta_\mu}{k}+\frac{p_\mu}{m^2(1-\frac{\eta p}{k})} \]\, xp ,\\
\label{ordpmu}
\hat{p}_\mu&=& p_\mu,
\end{eqnarray}
where $p_\mu\equiv -i\hbar\,\partial_\mu\,$. We would like to remark that Equations (\ref{ordxmu})and (\ref{ordpmu}) are new results obtained with the help of the GU formalism. 
Using the Equations (\ref{ordxmu})and (\ref{ordpmu}) and imposing the constraint (\ref{fi1dsr}), we obtain

\begin{eqnarray}
\label{ops}
\hat{x} \hat{p} &=& xp - \[ \frac{\eta p}{k}+ \frac{p^2}{m^2(1-\frac{\eta p}{k})}\] xp \nonumber \\ \nonumber \\& =& xp - \[ \frac{\eta p}{k}+ \frac{p^2(1-\frac{\eta p}{k})}{p^2}\] xp \nonumber \\ &=& 0.
\end{eqnarray}
Equation (\ref{ops}) shows that our solutions satisfy the second class constraints at the operator level. This important result indicates that our procedure is correct. Also, in the Eq.(\ref{ordxmu}) we can move the operator $p_\mu$ from the denominator to the numerator by performing a binomial expansion in powers of $\frac{\eta p}{k}$. Then the operator $\hat{x}_\mu$ can be written as

\begin{eqnarray}
\label{ordxmue}
\hat{x}_\mu=x_\mu -  \[\frac{\eta_\mu}{k}+\frac{p_\mu}{m^2}(1+\frac{\eta p}{k}-\(\frac{\eta p}{k}\)^2+\dots) \]\, xp .
\end{eqnarray}
We can observe that Eq.(\ref{ordxmu}) is equivalent to the series, Eq.(\ref{ordxmue}), with higher order terms in $p_\mu$.
We must mention the ordering problem that appears in Eq.(\ref{ordxmue}). Again, this problem can be solved by using the Weyl ordering operator prescription.

\section{Conclusions}
In this paper we have developed a procedure that can be used  to solve the Dirac bracket commutators. Our solution is based on the property that the GU variables satisfy the Dirac brackets algebra if we strongly impose the discarded second class constraint. In principle, to apply our formalism, it is necessary to know the exact form of the gauge invariant variables, Eq.(\ref{series}). The application of our formalism in systems with more than two second class constraints and the inclusion of fermions variables will be studied in future papers.

\section{Acknowledgments}
We would like to thank Andr\'{e}. G. Sim\~ao for critical reading.

\end{document}